**Modeling granular material blending in a rotating drum using a finite element method and advection-diffusion equation multi-scale model**

Yu Liu[a], Marcial Gonzalez[a,*] and Carl Wassgren[a,b]

[a]*School of Mechanical Engineering, 585 Purdue Mall, Purdue University, West Lafayette, IN 47907-2088, U.S.A.*

[b]*Department of Industrial and Physical Pharmacy (by courtesy), 575 Stadium Mall Drive, Purdue University, West Lafayette, IN 47907-2091, U.S.A.*

* Corresponding author at: School of Mechanical Engineering, Purdue University, West Lafayette, IN 47907-2088, U.S.A., Tel.: +1 765 494 0904,

*e-mail address*: marcial-gonzalez@purdue.edu (M. Gonzalez).

**Abstract**

A multi-scale model is presented for predicting the magnitude and rate of powder blending in a rotating drum blender. The model combines particle diffusion coefficient correlations from the literature with advective flow field information from blender finite element method simulations. The multi-scale model predictions for overall mixing and local concentration variance closely match results from discrete element method (DEM) simulations for a rotating drum, but take only hours to compute as opposed to taking days of computation time for the DEM simulations. Parametric studies were performed using the multi-scale model to investigate the influence of various parameters on mixing behavior. The multi-scale model is expected to be more amenable to predicting mixing in complex geometries and scale more efficiently to industrial-scale blenders than DEM simulations or analytical solutions.

*Topical Heading*: Particle Technology and Fluidization

**Introduction**

Blending of particulate materials is a common manufacturing unit operation in many industries, such as those that produce chemicals, pharmaceuticals, food products, and agrochemicals. Generating a homogeneous mixture can be critical to product quality and performance and, thus, proper design and operation of a blending operation is essential.[1]

Unfortunately, tools for quantitatively predicting particulate blending processes are lacking. Most often, parameters that produce an acceptable degree of blending are determined empirically. Numerical simulations using the discrete element method (DEM) have been used in recent years to predict blending unit operations, but these models must assume particle sizes orders of magnitude larger than the true particle sizes.[2-8] This assumption calls into question the quantitative accuracy of the predictions, particularly at smaller scales of scrutiny. In this paper, we present a new multi-scale modeling approach for predicting blending of a particulate material in a rotating cylindrical drum. A finite element method (FEM) simulation is used to predict the shape of the material domain and the bulk material velocity field. This information is used within a finite difference formulation of the advection-diffusion equation, which predicts the concentration evolution, i.e., blending, of the material. The diffusion coefficients used within the advection-diffusion equation come from the literature and were originally found both computationally and experimentally from small-scale, simple geometries. The model is considered multi-scale in the sense that a continuum model is used for bulk flow behavior while a different model is used to for particle diffusive behavior. A drum is considered since it is the simplest, tumbling blender geometry, but the modeling approach described in this paper can be used for more complex systems. Indeed, a significant advantage of the current modeling approach compared to previously published models is that arbitrary system geometries can be modeled. Moreover, this new modeling approach is well suited for predicting blending in industrial-scale systems, which are beyond the scope of current DEM modeling techniques.

**Background**

There has been considerable effort focused on developing dynamic models for predicting particle mixing and segregation with varied success.[9-13] Besides, computational models are also used and most of these computational studies involve the use of discrete element method (DEM) computer simulations. DEM simulations are particularly helpful for understanding blending physics at the particle level. For example, recent DEM simulations by Fan et al. and Khola and Wassgren produced expressions for particle diffusion coefficients and segregation rates using simple heap flow and shear cell geometries, respectively.[14,15] Other studies have investigated the effects of particle size, shape and cohesion, also in simple geometries.[16-19] Unfortunately, due to computational limitations, DEM simulations are not well suited for quantitative blending predictions at scales of industrial interest, at least using realistic particle sizes. Typical DEM simulations model on the order of $10^5$ particles at most, with some simulations reaching as many as $10^6$ particles. However, typical industrial blending operations involve $10^{12}$ particles. To maintain the same fill level in the model that is used in the real process, particle sizes are made artificially large in the DEM simulations, often by two to three orders of magnitude. Although DEM can still produce qualitatively reasonable mixing behavior using such large particles, it is not clear that the models are accurate quantitatively. Indeed, studies by Sarkar and Wassgren found that changing particle size affected the rate of axial mixing in a continuous paddle blender.[17]

In addition to issues related to particle size, the use of DEM simulations also requires knowledge of a number of particle material properties, such as elastic modulus and Poisson's ratio, and particle interaction properties, such as coefficients of restitution and sliding and rolling friction coefficients. Direct measurements of these properties are often impractical, particularly for particles smaller than 1 mm. Often, parametric studies are performed to determine the sensitivity of the simulation results on the unknown properties in order to account for parameter value uncertainty. Backfitting of bulk simulation results to bulk experimental measurements is becoming increasingly common, but questions remain as to whether (a) different sets of parameters might also fit the experimental results well, and (b) inaccuracies in the DEM model are disguised by parameter fitting.[20-22] Ideally, blending simulations would rely on a small number of well-defined, easy-to-perform bulk level measurements.

Recently, significant progress has been made toward addressing some of these issues. Zheng and Yu used the finite element method (FEM) assuming Mohr-Coulomb continuum material behavior to predict material velocity and stress fields in rotating drums and hoppers.[23-25] Other researchers have attempted similar FEM continuum modeling efforts for describing powder flow, but with varied success.[26-29] The advantages of assuming continuum material behavior are that (a) simulations at an industrial scale can be performed since tracking individual particles is not required, and (b) material characterization is straightforward, using standardized shear cell equipment for example.

Another significant recent modeling advance is the combination of particle-level blending models with macroscopic-scale flow fields. Several researchers have combined analytical expressions for particle diffusion and segregation at a local scale with analytical expressions for a macroscopic flow field.[30-32] These models have been used to gain good agreement with experiments.[33,34] In particular, previous work has developed a general theoretical framework for the quasi-two-dimensional granular flow in a rotating drum and provided simple models that give insight into how granular materials mix.[35,36] More recently, researchers combined correlations for particle diffusion and segregation at a local scale from DEM simulations with analytically-derived macroscopic flow fields.[16,37] The predictions from these studies have been accurate as well. The primary deficiency with this modeling approach is that the system geometries must be simple enough so that an analytical solution for the macroscopic flow field is available. Recently, Bertuola et al. used an FEM model to derive the flow field in a discharging hopper and combined it with the correlations for particle segregation by Fan et al. and Hajra et al.[16,38,39] Their FEM model treated the material as a non-Newtonian fluid with a local viscosity dependent on local particle granular temperature. They report that their model was able to quantitatively predict the degree of segregation at the discharge of the hopper measured in published experiments, but key model parameters needed to be fit to the experimental data for agreement to occur. Newly published work by Bai et al. predicts the flow field in a cylindrical, bladed mixer using an FEM model.[40] They used their model to qualitatively predict the mixing observed in DEM simulations, albeit with a dependence of the mixing rate on mesh size. This undesired dependency is attributed to an artificial diffusion intrinsic to the numerical method, i.e., as mesh size approaches zero, mixing is solely controlled by advection. Evidently, a physical advection-diffusion model is needed for a better and efficient quantitative prediction of the mixing process.

The current work investigates blending in a rotating drum using a multi-scale modeling approach similar to Bertuola et al.[38] However, several key implementation details are different. First, rather than predicting segregation in a discharging hopper, the current work investigates blending in a rotating drum. Since only blending is investigated, a correlation for particle segregation at the local scale is not needed, but an expression for diffusion coefficient is. We make use of correlations from the literature with no fitting parameters. Note that the term "diffusion" in the current work refers to Taylor dispersion of particles as opposed to true, Brownian diffusion. We retain the term diffusion, however, to remain consistent with prior studies. Second, the macroscopic velocity field is found via an FEM simulation assuming a Mohr-Coulomb material, similar to Zheng et al.'s work.[23-25] The Mohr-Coulomb properties are measured from simple, standard tests. It is important to emphasize that none of the multi-scale model parameters are back-fit to achieve good modeling accuracy, but rather are measured from independent tests or obtained from the literature. The multi-scale model is used to predict blending results from DEM simulations as a point of comparison. The current work introduces the finite element modeling approach and describes the advection-diffusion equation used in the multi-scale model and the numerical method used to solve it. It also describes the comparisons between the DEM simulations and multi-scale model predictions. Parametric studies are performed to help understand the effects of different parameters in the multi-scale model.

**Finite element method model**

A three-dimensional FEM model is used in the present work to provide predictions of the advective flow field in a cylindrical rotating drum. Prior works by Zheng and Yu have shown that FEM models can provide good predictions of the flow behavior of bulk granular materials.[23-

[24] The following sub-sections describe the FEM model implementations used in the present work.

*Model and boundary conditions*

The FEM model used here is derived from the one described by Zheng and Yu.[23] For convenience, several aspects of this model are presented here. The commercial FEM package Abaqus/Explicit V6.14 is used to perform the simulations. The system geometry is shown in Figure 1 for a lab scale rotating drum with a diameter of 140 mm. A narrow width of 1 cm is used for computational efficiency. Note that both the front and back sides of the Eulerian mesh are regarded as planes of symmetry in the model, which is analogous to having periodic boundaries. The drum wall is meshed separately as a rigid shell, with the only degree of freedom being rotation about the $z$ axis. The rotational speed remains constant throughout the simulations and equal to 6 rpm (0.628 rad/s). Gravity is included in the model with $g = 9.8$ m/s$^2$ directed in the negative $y$ direction.

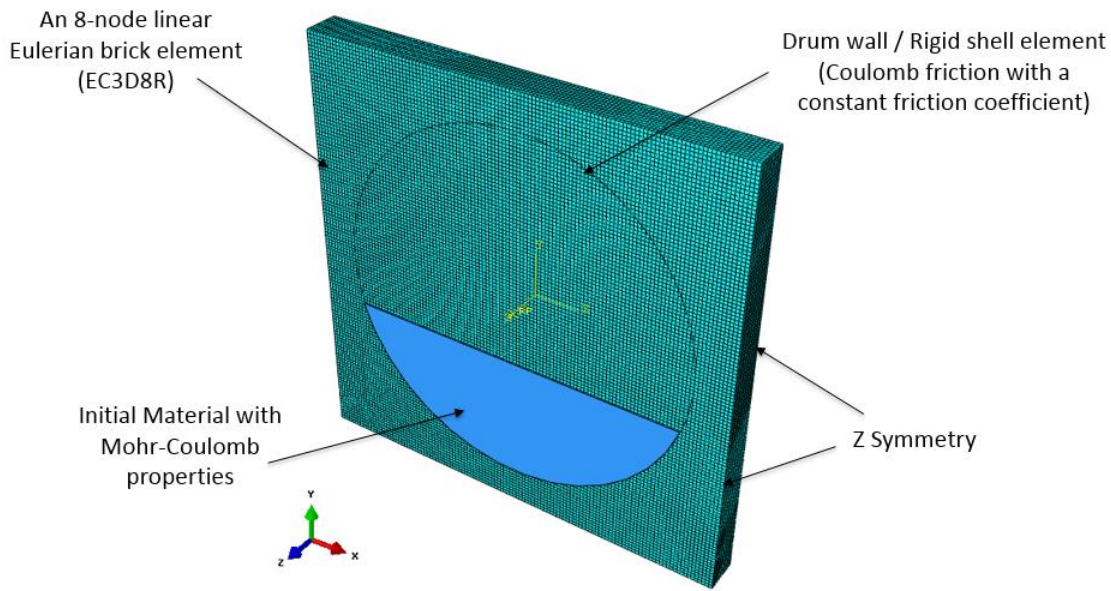

Figure 1. A schematic of the geometry modeled in the FEM simulations.

Because the material in the drum is anticipated to deform substantially, an Eulerian element formulation is used in the model to contend with potential mesh distortion issues. The entire simulation domain is modeled using 8-node, linear, Eulerian brick elements (EC3D8R) while the drum circumference is assumed to be a non-deformable, frictional, Lagrangian boundary. Within the Eulerian domain, the material stress-strain behavior is modeled using the Mohr-Coulomb elastoplastic (MCEP) model. Upon yielding, the material flows plastically. The continuity equation is discretized using a coupled Eulerian-Lagrangian formulation to efficiently model free surfaces and rigid walls. Details of this model can be found in Abaqus documentation.[41] Previous work has shown that the MCEP model can describe the behavior of flowing granular materials well.[23-25] Moreover, as reported by Zheng and Yu, the Young's modulus, Poisson's ratio, and bulk density of the material have little influence on the material flow behavior and, hence, can be treated as constant values.[23] Note that although the FEM algorithm includes inertial terms in the momentum equation, there is no stress dependence on strain rate in the MCEP material model. There are other constitutive models available that do include strain-rate effects.[42-44] The methodology for obtaining the MCEP material properties used within the simulation are described in the following sections.

*Abaqus implementation*

Since an Eulerian element formulation is used in the current model to eliminate mesh distortion, the Coupled Eulerian-Lagrangian (CEL) approach in Abaqus is applied to handle interactions between the highly deformable material elements and stiff wall elements. Details of this approach can be found in the Abaqus documentation.[41] In the Eulerian framework, elements do not represent specific masses of material, but instead represent specific regions in space. The volume of material within an element is represented by the Eulerian Volume Fraction (EVF). A value of EVF = 0 indicates that no material is present in the element while EVF = 1 indicates that the element is completely filled with material.

All Eulerian elements were initially empty (EVF = 0) and the initial bed state was generated by filling a fraction of the elements with material (EVF = 1), which is highlighted in blue in Figure 1. Next, the material was allowed to settle as the gravitational acceleration was slowly increased from zero to its final value. At this point, the drum was allowed to rotate and the simulation was considered started. This gravity-varying filling procedure was used to fill the drum and reduce the time needed for the material to completely settle before rotation. The filling level of the drum, defined as the maximum level bed depth divided by the drum diameter, was 0.35. Although the current work uses the FEM model to predict the macroscopic flow field in a rotating cylindrical drum, several trials were also performed to investigate the material stresses in the drum. Zheng and Yu showed that the CEL approach does not impose specific requirements on the shape or dimensions of the Eulerian element mesh since the mesh can cover the entire domain of the modeled system.[23] However, in the current work it was observed that the predicted material stresses are sensitive to the Eulerian domain shape. Figure 2 shows two different meshing schemes for the Eulerian domain. Figure 2(a) is the rectangular structured mesh used by Zheng and Yu while Figure 2(b) is a cylindrical structured mesh tested in the current work. Note that the central hole in Figure 2(b) is used to maintain the structured mesh for the cylindrical domain and does not affect the results since there is no material within that region. Mesh dependence tests were carried out for both meshing algorithms to verify that the meshes were sufficiently resolved for the moving system. As shown in Figure 3, the cylindrical structured mesh provides better predictions of the hydrostatic stress field, especially near the rigid drum wall, since this meshing geometry ensures that the rigid wall does not cut through Eulerian elements and allows for better contact detection. Nevertheless, since the current work does not rely on a predicted stress field, no further study was performed on this topic. As mentioned previously, the mesh geometry plays little role on the velocity field; hence, a rectangular structured mesh was used due to its computational efficiency.

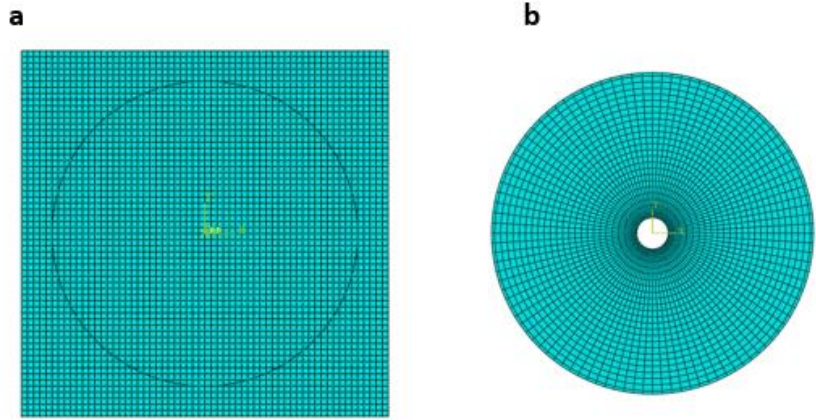

Figure 2. Two different meshing algorithms for the cylindrical drum (side view): (a) rectangular structured mesh, (b) cylindrical structured mesh, for the Eulerian domain.

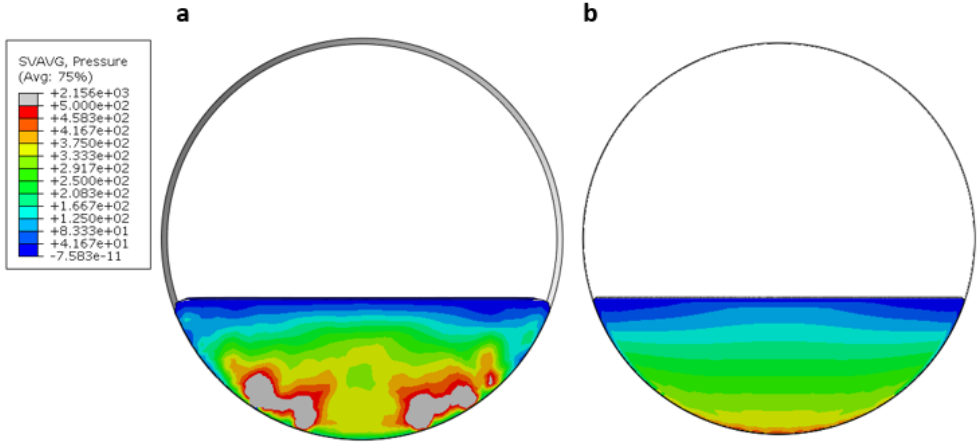

Figure 3. Stress distributions for the two meshes shown in Figure 2: (a) rectangular structured mesh, (b) cylindrical structured mesh. The colors represent pressure in Pascals.

**The multi-scale blending model**

*Advection-diffusion equation*

As mentioned previously, the FEM simulations provide only the velocity field information for material movement in the blender. Hence, in order to model the spatiotemporal evolution of the concentration of a particular material species, *c*, an additional model is needed. This model is the advection-diffusion equation,

$$\frac{\partial c}{\partial t} = \nabla \cdot (D \nabla c) - \nabla \cdot (\mathbf{v} c), \tag{1}$$

where *c* is the local concentration of a particular species of material, *D* is the diffusion coefficient for that species, and *v* is the local advective velocity vector. For simplicity, incompressible flow is assumed in the current work, which gives the local mass conservation equation,

$$\nabla \cdot \mathbf{v} = 0. \tag{2}$$

Previous studies have been devoted to understanding the underlying mechanisms governing particle mixing and developing analytical and numerical methods for predicting blending dynamics.[10,11,16] These modeling frameworks incorporated advective movement of material using either DEM simulation measurements or theoretical expressions for simple geometries, such as steady flow down a free surface. The primary difference with the present work is that the advective flow field is generated using an FEM model, which greatly increases the types and sizes of systems that can be modeled.

Since the current system of interest, a cylindrical rotating drum, is nominally two-dimensional, the advection-diffusion equation is simplified for a two-dimensional case. Note that there are no significant limitations to modeling fully three-dimensional flows, however. Since the current model only focuses on self-diffusion during blending, no segregation is considered. Future work will incorporate a segregation component into the multi-scale modeling framework.

Previous work has treated the self-diffusion coefficient $D$ as a constant value for simplicity.[10,16] However, studies have shown that $D$ is, in fact, a tensor quantity with components $D_{ij}$.[45-47] Utter and Behringer found that the off-diagonal components $D_{xy}$ and $D_{yx}$ are an order of magnitude smaller than the diagonal components $D_{xx}$ and $D_{yy}$ and, hence, can be reasonably ignored.[45] The advection-diffusion equation (Eq. (1)) written in index notation and making use of Eq. (2) is,

$$\frac{\partial c}{\partial t} = \frac{\partial}{\partial x_i}\left(D_{ij}\frac{\partial c}{\partial x_j}\right) - v_i\frac{\partial c}{\partial x_i}, \tag{3}$$

which may be expanded to give,

$$\frac{\partial c}{\partial t} = \left(\frac{\partial D_{xx}}{\partial x} - v_x\right)\frac{\partial c}{\partial x} + D_{xx}\frac{\partial^2 c}{\partial x^2} + \left(\frac{\partial D_{yy}}{\partial y} - v_y\right)\frac{\partial c}{\partial y} + D_{yy}\frac{\partial^2 c}{\partial y^2}, \tag{4}$$

taking into account the diffusion coefficient assumptions. Utter and Behringer also found that the particle diffusivity is proportional to the local shear rate and is approximately 1.9 times larger along the mean flow direction than it is in the perpendicular direction.[45] A similar relationship has been reported by Hsiau and Shieh.[46] More recently, Fan et al. measured the self-diffusion coefficient $D$ in the spanwise direction of a heap flow and found that when the shear rate is not too small,[14]

$$D \sim \dot{\gamma}\bar{d}^2, \tag{5}$$

where $\dot{\gamma}$ is the local shear rate in the direction perpendicular to bed free surface and $\bar{d}$ is the local mean particle diameter.

Since both the local shear rate $\dot{\gamma}$ and diffusion coefficient $D$ are tensors, they can be decomposed into x and y components in order to solve Eq. (5) in a structured mesh. Combining decomposed

Eq. (5) with Utter and Behringer's relationship, the shear rate-dependent diffusion coefficient $D$ can be written as,

$$D_{xx} = k_1 \dot{\gamma}_y \bar{d}^2 + k_2 \dot{\gamma}_x \bar{d}^2$$
$$D_{yy} = k_1 \dot{\gamma}_x \bar{d}^2 + k_2 \dot{\gamma}_y \bar{d}^2. \quad (6)$$

where $\dot{\gamma}_y = |\partial v_y / \partial x|$ and $\dot{\gamma}_x = |\partial v_x / \partial y|$. The constant $k_1$ can be found from experiments or small-scale DEM simulations. In the current work, $k_1 = 0.04$, which is derived from the previous computational work by Fan et al.[14] Making use of Utter and Behringer's findings, $k_2 = 1.9 k_1$.[45]

*Numerical method*

Many numerical techniques have been developed to solve the advection diffusion equation, such as the finite difference method, finite element method, finite volume method, and the domain decomposition method. Previous works also introduced an operator splitting approach, which solves the advection step and a combined diffusion and segregation step separately.[10,11,16] A matrix mapping method is used in these previous works due to its high accuracy.[48] Since the current model is two-dimensional and only self-diffusion is considered, a finite difference method using a central explicit scheme is used here to solve Eq. (4) due to its simplicity and computational efficiency.

Previous work has shown that a simple combination of individual finite differences in different axes without appropriate cross terms can cause numerical instability.[49] The generalization of explicit finite difference schemes for the advection-diffusion equation to multiple dimensions is not simply the sum of individual one-dimensional contributions. Hence, in order to maintain a high computational efficiency without losing accuracy, a second-order Tylor Lax-Wendroff scheme is used in the current model to rewrite Eq. (4) as,

$$c_{ij}^{n+1} = c_{ij}^n - \left[ v_x \Delta_{x0} c_{ij}^n - \left( \frac{1}{2} v_x^2 + \mu_x \right) \delta_x^2 c_{ij}^n \right] - \left[ v_y \Delta_{y0} c_{ij}^n - \left( \frac{1}{2} v_y^2 + \mu_y \right) \delta_y^2 c_{ij}^n \right] + v_x v_y \Delta_{x0} \Delta_{y0} c_{ij}^n, \quad (7)$$

where,

$$v_x = \frac{\left[ v_x + \Delta_{x0} (D_{xx})_{ij} \right] \Delta t}{\Delta x}, \quad (8)$$

$$v_y = \frac{\left[ v_y + \Delta_{y0} (D_{yy})_{ij} \right] \Delta t}{\Delta y}, \quad (9)$$

$$\mu_x = (D_{xx})_{ij} \frac{\Delta t}{\Delta x^2}, \quad (10)$$

$$\mu_y = (D_{yy})_{ij} \frac{\Delta t}{\Delta y^2}, \quad (11)$$

$$\Delta_{x0} c_{ij}^n = \frac{c_{i,j+1}^n - c_{i,j-1}^n}{2}, \quad (12)$$

$$\Delta_{y0} c_{ij}^n = \frac{c_{i+1,j}^n - c_{i-1,j}^n}{2}, \quad (13)$$

$$\delta_x^2 c_{ij}^n = c_{i,j+1}^n - 2c_{ij}^n + c_{i,j-1}^n, \quad (14)$$

$$\delta_y^2 c_{ij}^n = c_{i+1,j}^n - 2c_{ij}^n + c_{i-1,j}^n, \quad (15)$$

$$\Delta_{x0} \Delta_{y0} c_{ij}^n = \frac{\left( c_{i+1,j+1}^n - c_{i-1,j+1}^n - c_{i+1,j-1}^n + c_{i-1,j-1}^n \right)}{4}, \quad (16)$$

$$\Delta_{x0} (D_{xx})_{ij} = \frac{(D_{xx})_{i,j+1} - (D_{xx})_{i,j-1}}{2}, \text{ and} \quad (17)$$

$$\Delta_{y0} (D_{yy})_{ij} = \frac{(D_{yy})_{i+1,j} - (D_{yy})_{i-1,j}}{2}. \quad (18)$$

This finite difference formula is illustrated using the computational molecule shown in Figure 4, which can be used independently of the directions of the velocity components. Details of this scheme, including its Taylor series expansion and the iteration process, can be found in the work by Sousa.[49]

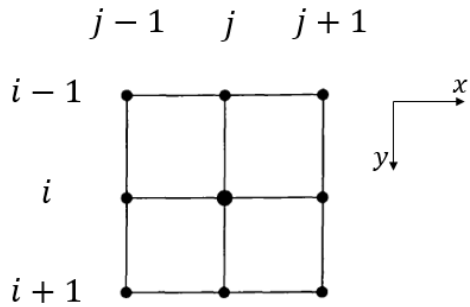

Figure 4. Computational molecule for the second-order Taylor Lax-Wendroff scheme. The larger circle denotes the center node.

Note that since an explicit finite difference scheme is used here, solutions of the current model are not unconditionally stable. Hence, the Von Neumann conditions must be checked before the iteration to ensure the stability of the numerical computations. Details of this stability analysis can be found in the published literature.[49,50] However, in practice, an easier way to ensure numerical stability is to continuously decrease the time step until reaching a critical value below which the numerical computation remains stable. Using this approach, the time step in the current model is determined as 5e-4 s (20,000 time steps per drum rotation) with the grid length of 3e-4 m (467 elements span the drum diameter).

To determine the initial particle concentrations used in the multi-scale model, the material region in the FEM model was divided into two equal parts by assigning two different material colors as shown in Figure 5(a). All of the materials had the same properties, however. The drum was then allowed to move in an unsteady fashion according to the FEM predictions as the drum started to rotate and the material advected until a steady velocity field was achieved, which occurred after 0.25 revolutions (Figure 5(b)). Particle diffusion was not considered during this stage of the model. This assumption is addressed further in the following sections. Once a steady velocity field was established in the FEM model, the velocity components were then used within the advection-diffusion equation (Eq. (4)) to model material mixing.

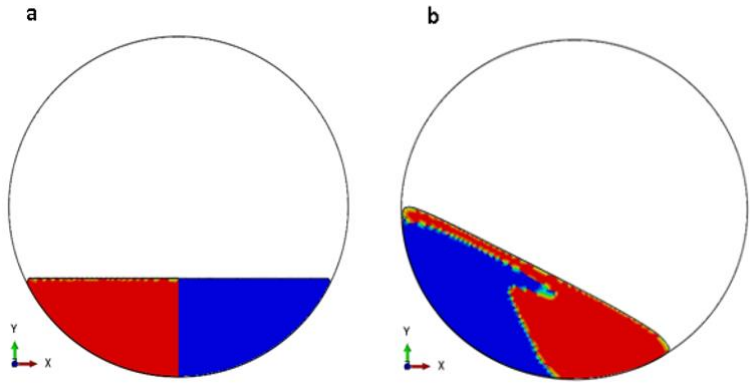

Figure 5. Material concentration distributions from a typical FEM simulation (a) initial filling and (b) after reaching steady state.

Because a steady state velocity field was used in the current model, the boundaries of the material domain remained the same throughout the entire computation. Hence, a control volume boundary condition was used in the current model and no moving boundary was considered. Note that in computational fluid dynamics (CFD) simulations, a polynomial fitting approach is often used and the species concentration of the boundary node is set equal to the extrapolated value of inner nodes of the control volume.[51] The same idea was used in the current algorithm and the material concentration of the boundary node was set to equal the value of the node that was one grid point inward.

Post-processing of the FEM simulation data was required to determine the extent of the material domain. As mentioned previously, the computational mesh in an Eulerian FEM does not represent the material and instead, the Eulerian Volume Fraction (EVF) is used to determine the volume and position of materials within the Eulerian mesh. Hence, nodal EVF values for the two materials must be generated first so that empty elements and boundaries can be identified.

A MATLAB program was used to iterate the finite difference form of the advection-diffusion equation given in Eq. (7). After obtaining the material domain and steady velocity field information from the FEM simulation, initial conditions were interpolated into a structured rectangular mesh that covered the entire material domain. The grid width was determined such that it was sufficiently small to achieve a converged result. Iteration was used for all interpolated nodes and a threshold was set to ensure the material concentration value remained between 0 and 1. As described previously, a small time step was carefully chosen to ensure the stability of the explicit scheme.

**Discrete element method (DEM) simulations**

*DEM rotating drum simulation*

Although DEM simulation is not well suited for industrial-scale blenders, it can still accurately predict the mixing and segregation for large particles.[2-8] Thus, a three-dimensional DEM model with large particles was developed in the current work to compare to the predictions from the multi-scale model. The commercial DEM package EDEM$^{TM}$ (DEM Solutions, Inc., Lebanon, NH) was used to perform the simulation. The system geometry is shown in Figure 6(a). Note that the front and back boundaries were periodic, consistent with the FEM model although the FEM symmetric boundary condition did not allow for material movement through the boundary while the DEM periodic boundaries did. Note that previous work has shown that sidewall friction can play an important role in the surface flow dynamics.[52] In the simulation, a drum was filled with two different colored, but otherwise identical, spherical particles, which were initially separated side by side as shown in Figure 6(b). Once the bed was fully settled, the drum rotated at a constant speed and the particle began mixing. The DEM parameters used in the simulation are listed in Table 1.

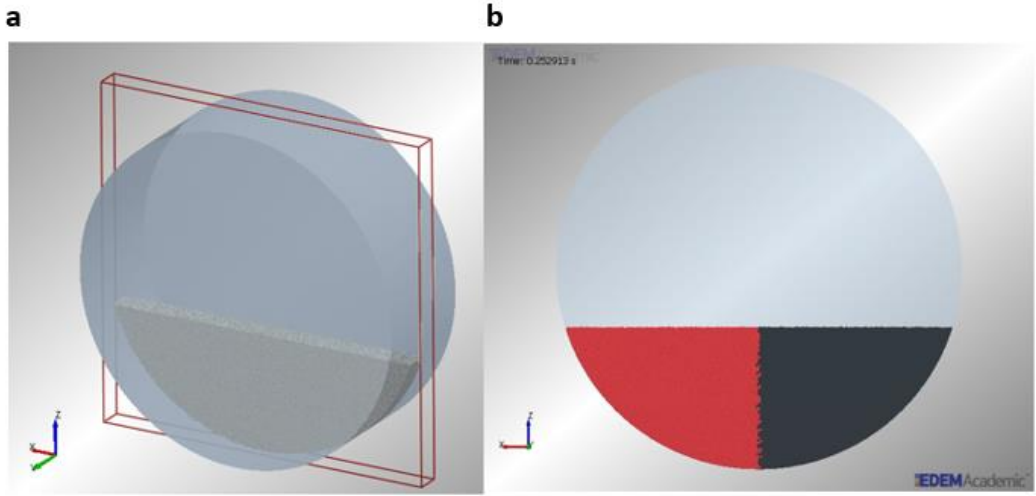

Figure 6. DEM simulation: (a) drum and domain geometry; (b) initial state.

Table 1. DEM simulation parameters.

| Parameter | Value |
| --- | --- |
| Drum diameter (mm) | 140 |
| Domain width (mm) | 10 |
| Number of particles (-) | 53445 |
| Particle diameter (mm) | 1 |
| Particle density (kg/m$^3$) | 2500 |
| Particle shear modulus (MPa) | 10 |
| Coefficient of restitution | 0.8 |
| Poisson's ratio (-) | 0.3 |
| Particle-particle friction coefficient (-) | 0.4 |
| Particle-wall friction coefficient (-) | 0.3 |
| Rolling friction coefficient (-) | 0.01 |
| Filling level (% of max level depth) | 35 |
| Rotation speed (rpm) | 6 |
| Simulated time (s) | 30 |
| Number of processors (-) | 16 |
| Wall-clock time (days) | 5 - 6 |

*Calibration of material properties*

In order to justify a comparison between the multi-scale model and the DEM simulations, the Mohr-Coulomb properties used in the FEM simulations were determined from the DEM particle properties specified in Table 1. The Mohr-Coulomb properties, namely the internal friction angle, cohesion, and dilation angle, were calibrated using a DEM simulation of an annular shear cell with periodic boundaries (Figure 7(a)). Analogous to a real annular shear tester, vertical fins were attached to the top and bottom plates to ensure failure within the material. A constant normal pressure was applied to the top plate while the bottom plate moved with a constant rotation speed and the tangential shear stress was recorded and evaluated to determine the bulk internal friction angle after a constant stress level was reached.

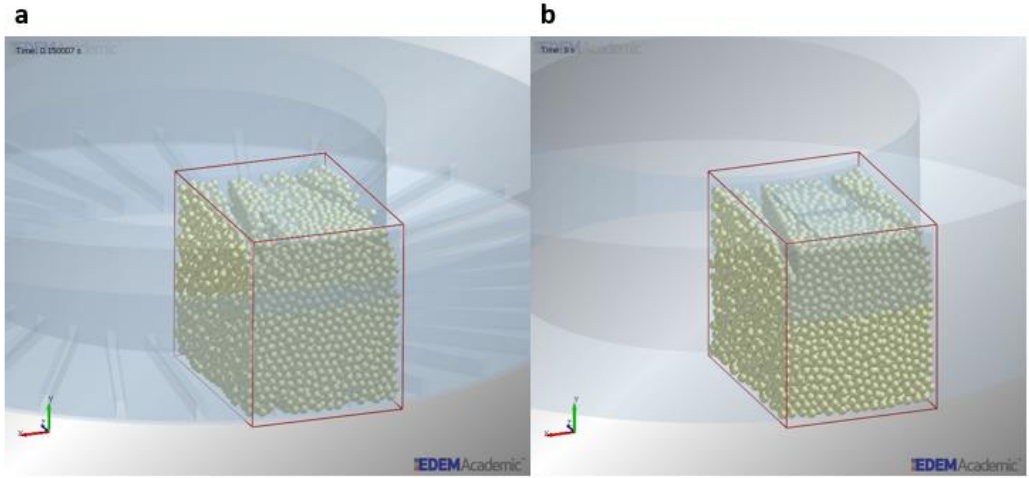

Figure 7. A snapshot showing the computational domain for the DEM annular shear cell simulations for (a) internal friction angle and (b) wall friction angle. Periodic boundary conditions are used in the $x$ direction.

The defining feature of a Mohr-Coulomb material is its shear failure criterion,

$$\tau = c + \sigma \cdot \tan(\varphi), \tag{19}$$

where $\tau$ and $\sigma$ are the failure shear stress and the applied normal stress acting on the failure plane, respectively. The parameter $c$ is the bulk cohesion of the material and the parameter $\varphi$ is the internal friction angle of the material. These parameters were found using two different shear cell simulations with different top plate normal pressures: one at 2 kPa and the other at 4 kPa. Figure 8 plots the results from these two simulations along with a best fit line based on a least squares fit to Eq. (19). The results clearly demonstrate that the bulk material can be treated as cohesionless ($c = 0$) with an internal friction angle of 23.6°. The dilation angle of material, $\psi$, was set to 0.1° since the dilation of cohesionless granular materials is usually small, as stated by Zheng and Yu.[23] A wall friction simulation was also performed using a similar DEM annular shear cell, but with a flat bottom boundary surface, to determine the material-wall friction angle.

This simulation gave a bulk material-wall friction coefficient of 0.324, which is close to the particle-wall friction coefficient.

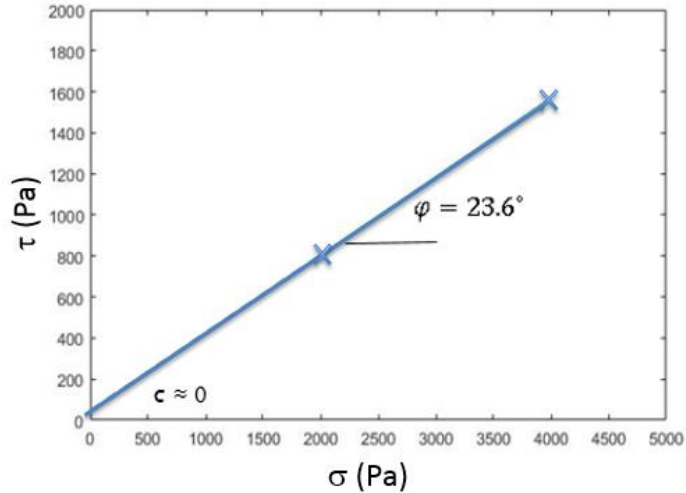

Figure 8. Critical state shear stress plotted as a function of the applied normal stress from the DEM shear cell simulations using the material properties listed in Table 1. Two data points are shown in the plot along with a fitting line.

Although previous work has indicated that Mohr-Coulomb FEM simulations of bulk material flow is insensitive to the elastic modulus, Poisson's ratio, and bulk density of the material, they were still measured from the DEM simulations in the current work for completeness.[23] The bulk density was taken to be 0.6 of the particle density, which is consistent with loosely packed, non-cohesive spheres. The bulk elastic parameters, namely, elastic modulus $E$ and Poisson's ratio $\upsilon$, were obtained from a separate uniaxial compression DEM simulation in which the axial and radial stresses were measured as a function of axial strain during both compression and decompression of the material, as shown in Figure 9. The initial slope of the unloading curves were used to compute both parameters,

$$\upsilon = \frac{\frac{d\sigma_{rr}}{d\varepsilon^e_{zz}}}{\frac{d\sigma_{rr}}{d\varepsilon^e_{zz}} + \frac{d\sigma_{zz}}{d\varepsilon^e_{zz}}}, \quad (20)$$

$$E = \frac{d\sigma_{zz}}{d\varepsilon^e_{zz}} - 2\upsilon\frac{d\sigma_{rr}}{d\varepsilon^e_{zz}}, \quad (21)$$

where $\sigma_{zz}$ is the axial stress, $\sigma_{rr}$ is the radial stress, and $\varepsilon^e_{zz}$ is the elastic axial strain. Details of this method can be found in the work of Swaminathan et al.[53]

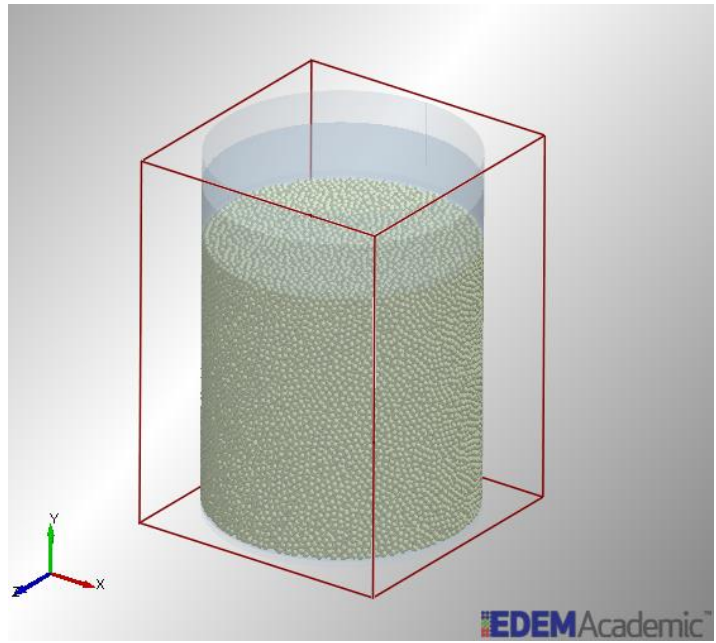

Figure 9. A snapshot showing the computational domain for the DEM uniaxial compression simulations.

Figure 10 shows the compression and decompression curves from the DEM simulation using the properties listed in Table 1. Previous work has shown that the bulk elastic properties are functions of the relative density, which in turn is a function of the applied stress.[53-55] From the FEM simulation, the stresses acting on material vary throughout the drum. Several DEM uniaxial compression simulations with maximum applied stresses between 20 kPa and 40 kPa have been performed and give values of (elastic modulus, Poisson's ratio) between (2.8 MPa, 0.055) and (3.7 MPa, 0.065). For simplicity, the elastic properties used in the current FEM model were determined for a single stress condition corresponding to Figure 10. The bulk elastic modulus $E$ and Poisson's ratio $\upsilon$ determined for this condition were 3.65 MPa and 0.065, respectively. Previous work has shown that the Poisson's ratio $\upsilon$ is typically small at relative densities near the poured relative density for a variety of different materials, which is consistent with the value measured in the current work.[53-55] As mentioned previously, the bulk material flow behavior is insensitive to the elastic modulus and Poisson's ratio so the variation in elastic properties is expected have little influence on the results. Indeed, FEM simulations performed as part of this work with different elastic properties produced little variation in the bulk material kinematics. A summary of the FEM simulation material parameters is given in Table 2. The drum geometry and rotation speed were identical between the two models. It is important to emphasize that the FEM model parameters in Table 2 were found from independent, standard material tests rather than being back-fit to the blending data. Since the present work compares the multi-scale model blending performance to results from a DEM simulation, DEM simulations were used to determine the FEM material parameters for consistency. For a more practical case, however, the FEM material parameters would be found from the same characterization tests (i.e., shear cell and uniaxial compression) but performed experimentally. Although spherical particles are used in the current work for simplicity, particle shape and size effects can be considered indirectly via the bulk material properties and diffusion coefficient used in the model. The diffusion coefficient could be measured experimentally or possibly found via DEM simulation.[14]

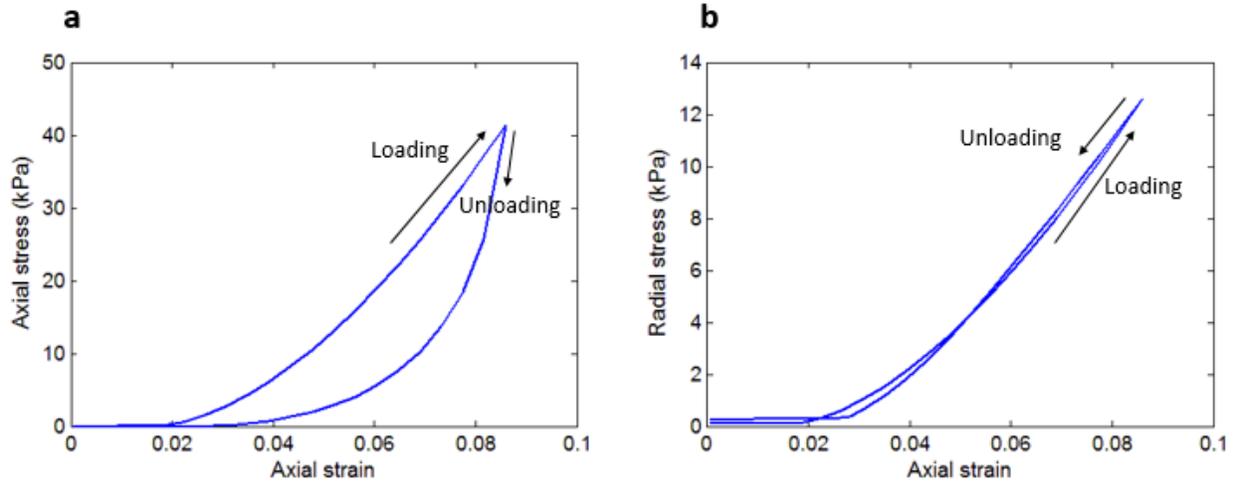

Figure 10. The compression and decompression curves from the uniaxial compression DEM simulation used to obtain the bulk elastic modulus and Poisson's ratio.

Table 2. Parameters used in the FEM simulation.

| Parameter | Value |
| --- | --- |
| Material density (kg/m$^3$) | 1500 |
| Young's modulus (MPa) | 3.65 |
| Poisson's ratio (-) | 0.065 |
| Internal friction angle (degree) | 23.6 |
| Cohesion (Pa) | 0 |
| Dilation angle (degree) | 0.1 |
| Wall friction coefficient (-) | 0.324 |

**Comparison of the DEM and multi-scale model results**

The magnitude of the steady state material velocity in the drum, predicted by DEM and FEM simulations, is shown in Figure 11. Figure 12 plots the velocity along the free surface of the material as well as in the surface-normal direction, predicted by the two different models. These two figures demonstrate that the FEM model compares favorably with the DEM models. The free surface angles predicted by the DEM and FEM simulations are 24.5° and 23.3°, respectively, with a difference of less than 5%. Moreover, the thicknesses of the active layer predicted by the two models differ by less than 5% as well. Zheng and Yu showed that their FEM rotating drum predictions matched experimental measurements well, lending confidence that the FEM model is a good model for predicting velocity fields.[23] Figure 11 also illustrates the well-known observation that the flow field in the drum can be divided into two distinct regions: an active region characterized by a thin, downward flowing layer adjacent to the free surface and a passive region below the active region where particles move in solid body rotation.[1] The large velocity gradient in the active region is the source of most of the diffusive mixing in the system (refer to Eq. (5)), which is consistent with previous work.[56]

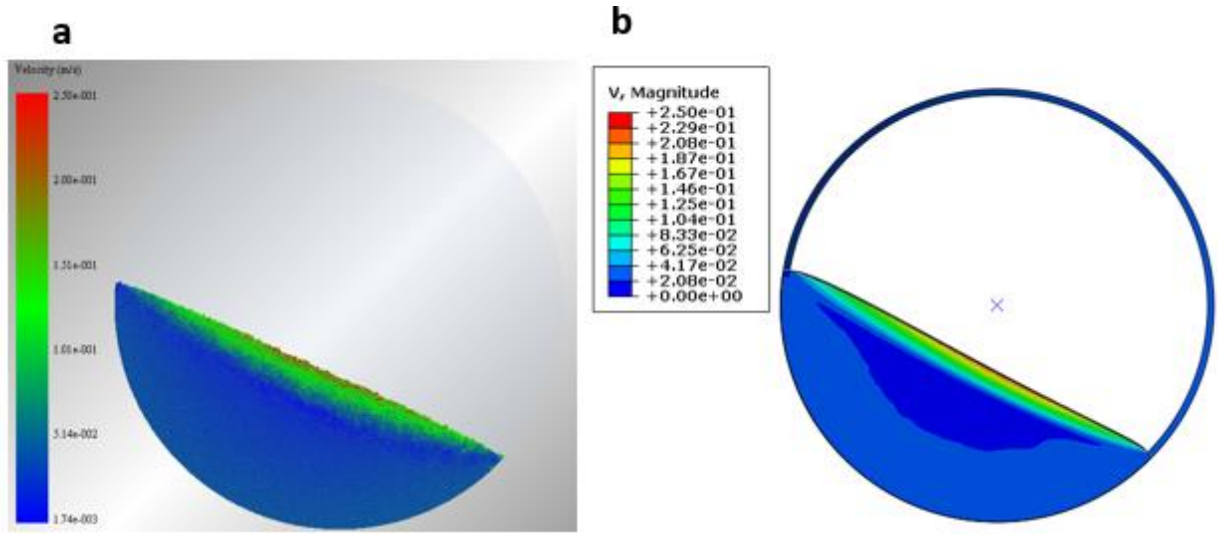

Figure 11. Field plots of the material speed in the (a) DEM simulation and (b) FEM simulation. The color scales used in the two figures are identical.

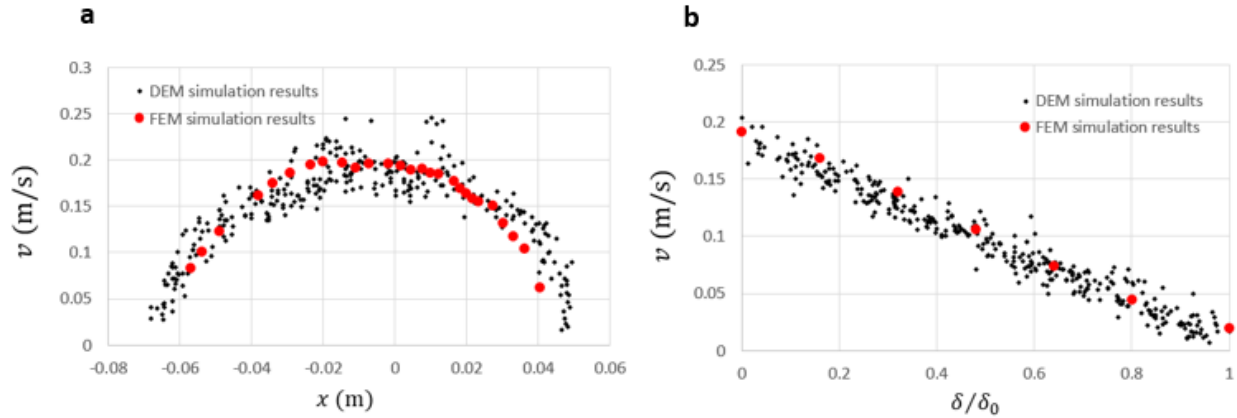

Figure 12. The velocity (a) along the free surface of the material, and (b) in the surface-normal direction in Figure 11, predicted by DEM and FEM simulations.

The initial stages of mixing are shown in Figure 13. To enforce consistency between the multi-scale model and DEM model, all the revolutions shown in the current work start from the horizontal position. This figure shows the state of the material in the DEM and FEM (advection only) models at 0.25 revolutions, after which the changes in the velocities in various regions of the drum were observed to be typically less than 3% and steady state material movement began. Clearly the mixing during this short unsteady period was dominated by material advection, which is consistent with the assumption described before that no diffusion happens during the initial development of the bed movement. As mentioned previously, the material concentration field predicted from the FEM model at this point was used as the initial condition for the multi-scale model. Note in Figure 13, the edges of the bed free surface in the FEM simulation show a rounder shape than in the DEM simulation due to the method Abaqus uses to display partially filled elements. This effect has no influence on the predictions of interest, namely the free surface angle, flow velocities, and mixing rate.

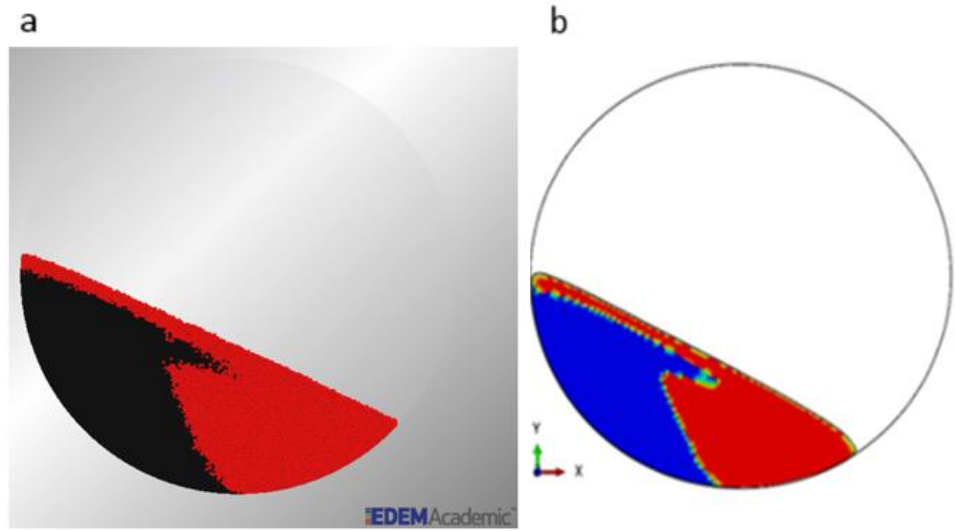

Figure 13. The material state in the (a) DEM simulation and (b) FEM simulation (advective material movement only) at 0.25 revolutions, which is when a steady state velocity field is reached.

The state of the material after different numbers of drum revolutions is shown in Figure 14 for both the DEM and multi-scale models. The colors in the DEM simulations are the individual red and black particles while the colors in the multi-scale model correspond to the concentration of red particles, with yellow indicating a large concentration of red particles and blue indicating a small concentration. As expected, as time increases the degree of mixing increases, with both advection and diffusion contributing to the mixing process as indicated by multi-scale model concentration values between zero and one (pure advection would only have concentration values of zero and one). At least qualitatively, the multi-scale model reproduces the mixing observed in the DEM simulations.

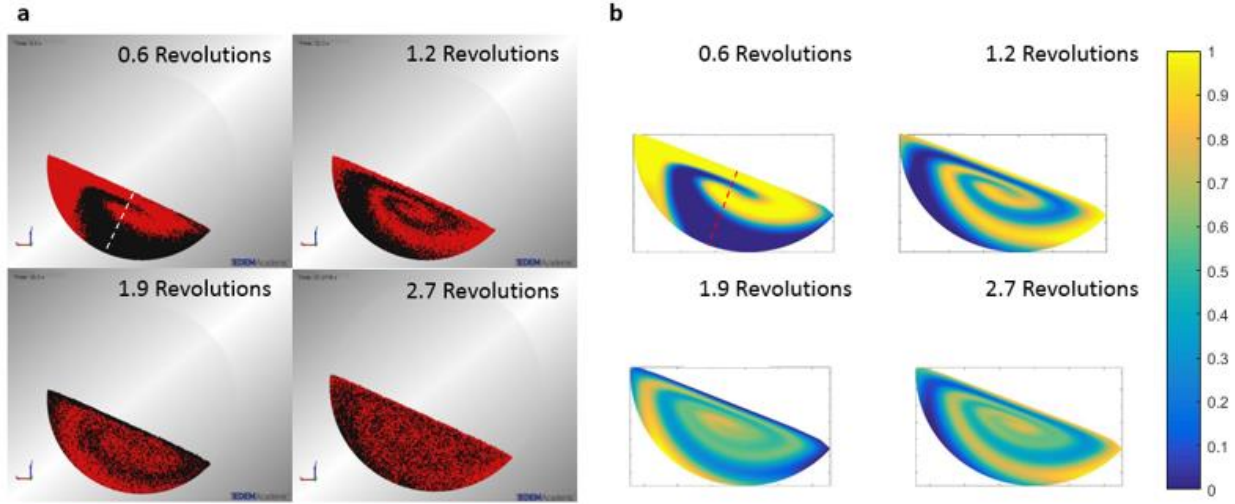

Figure 14. Snapshots showing the time evolutions of mixing. (a) DEM simulation and (b) multi-scale model. The vertical color scale in (b) is the red particle concentration.

To provide a more quantitative comparison of the two models, red particle concentration is plotted as a function of the dimensionless perpendicular distance $\delta$ along the center of the bed starting from the free surface (refer to the dashed lines shown in Figure 14). The distance is made dimensionless by dividing by the maximum level bed depth of the bed $h$. Thus, 0 (free surface) $\leq$ $\delta/h \leq 1$ (drum surface). In the DEM simulation, red particle concentration is calculated in cells with a square cross section and a depth spanning the drum width (10 particle diameters). The cells overlap, with each cell center located three particle diameters from its neighbors. The number of red and black particles with centers located within a cell are recorded to calculate the red particle concentration.

Note that because of the finite cell size, often referred to as the "scale of scrutiny", the concentration profile will vary with the cell size. Larger cell sizes provide less spatial resolution while cell sizes approaching the particle size produce less meaningful concentration values due to statistical fluctuations. To examine this cell size influence, three cell sizes of $3d$, $5d$, and $7d$ ($d$ is the particle diameter) are used in the concentration calculations. In the multi-scale model, there is no cell size since the material concentration is calculated directly from the advection-diffusion equation at each node point along the path.

Figure 15 plots the red particle concentration profiles for the DEM (with cell size of 5$d$) and multi-scale models for the same number of revolutions shown in Figure 14. There is very good quantitative agreement between the two models, although it does appear that there is a slight offset in the multi-scale model concentration values. As indicated previously, there is a minor difference in the free surface angle between two models, which will lead to an offset in the two results. Both sets of results show large peaks and valleys within the first two drum revolutions. These peaks and valleys diminish considerably by 2.7 revolutions due to increased advective folding of the material along with diffusion. With larger numbers of revolutions, the red particle concentration approaches the expected value of 0.5 along the entire profile. As expected, changing the cell size used in the DEM concentration calculation slightly affects the results. However, the difference is within 10% for a cell size of 5$d$ plus or minus 2$d$.

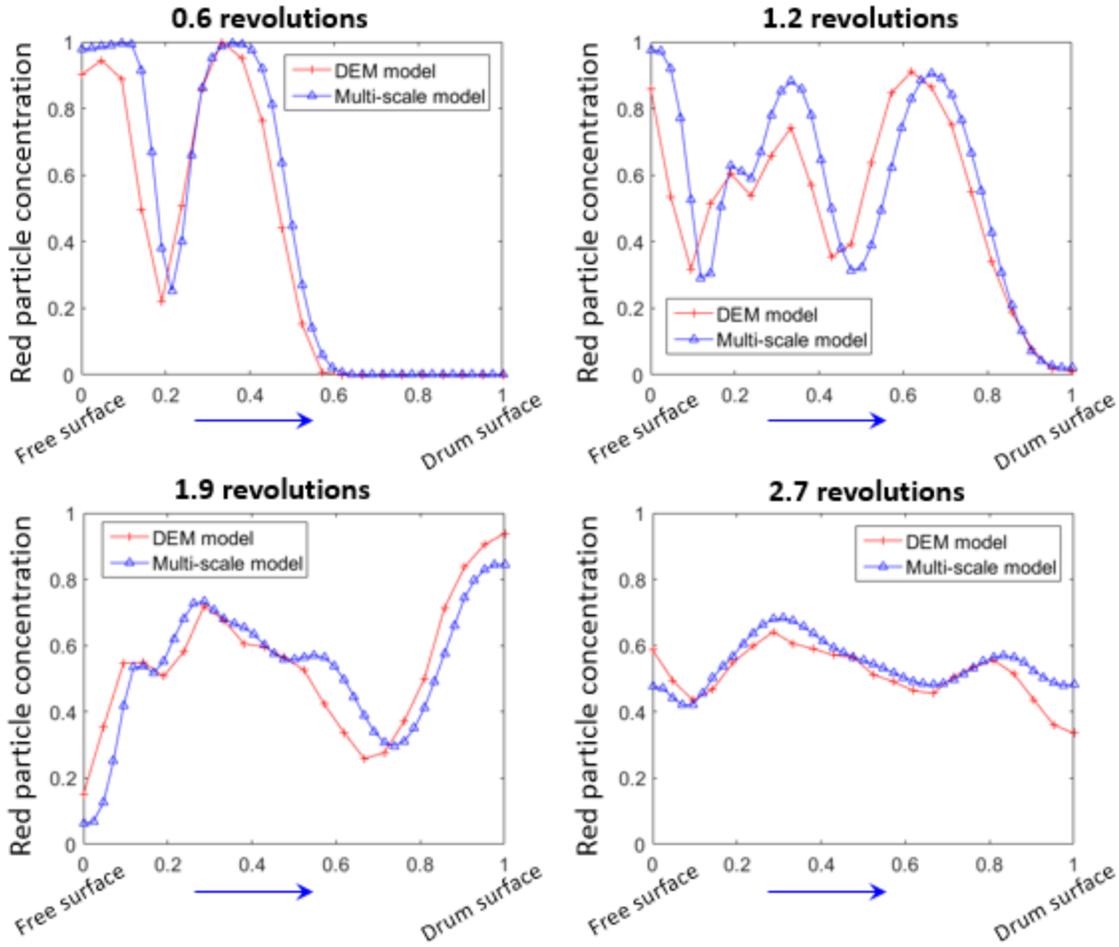

Figure 15. Red particle concentration plotted as a function of dimensionless distance from the free surface, $\delta/h$, along the centerline of the drum for both the DEM and multi-scale models. Each plot corresponds to a different number of drum revolutions.

An important point to make is that the multi-scale modeling approach is much faster than the DEM modeling approach. For the parameters listed in Table 1, the DEM simulation completed in 5 to 6 days of wall clock time using a 16 processor desktop PC. The same geometry and simulated time using the multi-scale modeling approach (Table 2 for the FEM portion) completed in 4 to 5 hours on the same PC with the same number of processors. The FEM computations comprised about 60% of this time with the MATLAB advection-diffusion calculations using the remainder. The multi-scale modeling approach is expected to be even more computationally efficient as the system size increases.

Another common measure used to assess the state of mixedness in a blending operation is the segregation intensity $I$, which is defined as,[57]

$$I_i = \frac{\sigma_i^2}{\sigma_0^2},  \tag{22}$$

$$\sigma_i^2 = \frac{1}{M-1}\sum_{m=1}^{M}(c_i - c_{i,m})^2,  \tag{23}$$

$$\sigma_0^2 = c_i(1-c_i),  \tag{24}$$

$$c_i = \frac{1}{M}\sum_{m=1}^{M}c_{i,m}.  \tag{25}$$

In these relations, $\sigma_i^2$ is the measured variance of component $i$'s concentration (here component $i$ is the red particles), $\sigma_0^2$ is the variance of component $i$'s concentration for a fully segregated system, $c_i$ is the measured mean concentration of component $i$, and $M$ is the total number of samples used to calculate the mean and variance. The segregation intensity varies from zero, corresponding to perfect mixing, to one, which is a fully segregated state. The segregation intensity provides less information and is a less stringent test of the multi-scale model's accuracy than the concentration profiles shown in Figure 15. Nevertheless, it is calculated here for both models since it is a common metric for assessing the state of blending.

In the multi-scale model, every node at which a concentration is calculated is used in the evaluation of the segregation intensity. For the DEM simulations, a non-overlapping grid of cuboidal cells, illustrated in Figure 16a, is used to calculate red particle concentration. As with the concentration profile, cell sizes of $3d$, $5d$, and $7d$ are used. The segregation intensity values are plotted in Figure 16b as a function of the number of drum revolutions. Both models display the frequently observed decay in segregation intensity, which is often fit to an exponential function.[58] Significantly, the multi-scale model quantitatively predicts the DEM results well. The DEM scale of scrutiny does play a minor role with larger scales of scrutiny having smaller segregation intensities. This trend is expected from statistics.[59]

The asymptotic values for the segregation intensity can be predicted analytically as originally described by Danckwerts.[59] A perfectly mixed state where $I = 0$ is generally not achievably in practice and instead a randomly mixed state is the expected asymptotic state. For a randomly mixed system, the concentration variance is,

$$\sigma_r^2 = \frac{c_i(1-c_i)}{N}, \qquad (26)$$

where $N$ is the number of particles in the cell used to calculate the concentration. Thus, the segregation intensity for a randomly mixed system is $I_R = \sigma_R^2/\sigma_0^2$. Clearly, as the scale of scrutiny increases, the asymptotic segregation intensity decreases, which is the trend observed in Figure 16b. Interestingly, the multi-scale model predicts an asymptotic segregation intensity of zero, corresponding to a perfectly mixed state. This result stems from the fact that the multi-scale model assumes the material is a continuum and thus the number of particles in Eq. (26) is infinite. Hence, although the multi-scale model accurately predicts mixing behavior throughout most of the process, it will always predict a smaller segregation intensity at larger times.

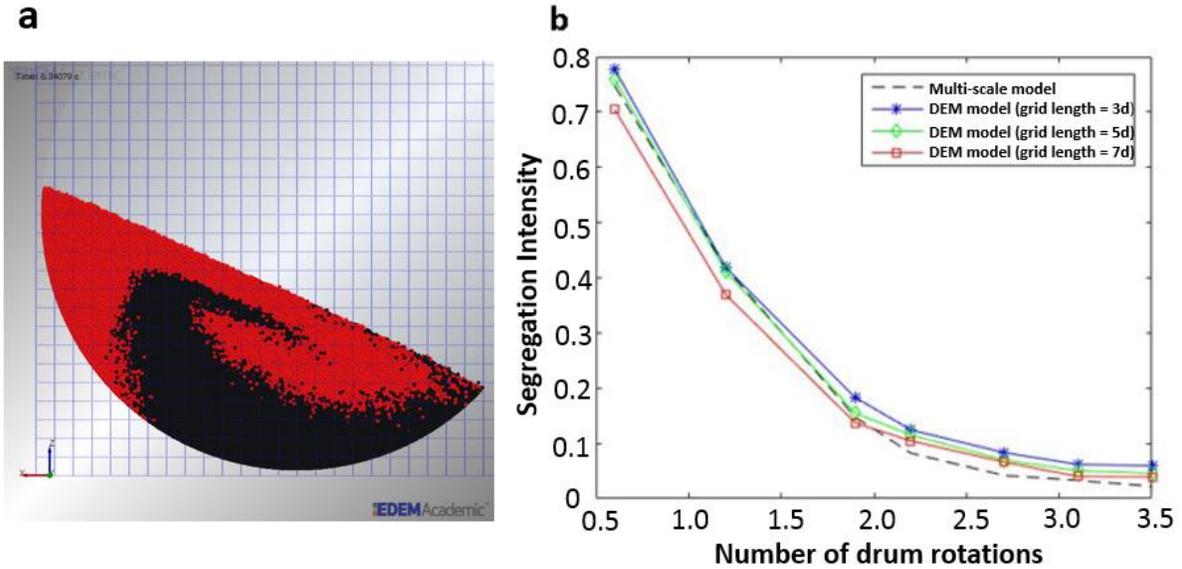

Figure 16. (a) The cuboidal cells in the DEM simulation; (b) the segregation intensity as a function of the number of drum revolutions for the DEM model and multi-scale model.

**Parametric study of the multi-scale model**

To better understand the effect of and the sensitivity to changes in different parameters of the multi-scale model, a study was performed in which several of the model parameters were varied. A sensitivity analysis of the system was also performed in order to more efficiently perform the simulations.

A mesh dependency study was performed first to ensure convergence of the advection-diffusion solution. Figure 17 shows the segregation intensity for different diffusion constants as a function of the number of drum revolutions for different numbers of mesh elements. For $k_1 = 0.04$, which was chosen based on prior studies, the solution was insensitive to the mesh sizes tested and, thus, 300,000 elements were used in the current study to maintain accuracy and computational efficiency (Figure 17(a)). Note that for a much smaller diffusion constant of $k_1 = 0.005$, a similar convergence analysis concludes that a mesh size of 600,000 is needed (see Figure 17(b)). This result indicates that for convection dominated flows, a refined mesh is required to avoid the artificial diffusion intrinsic to the finite difference method itself.

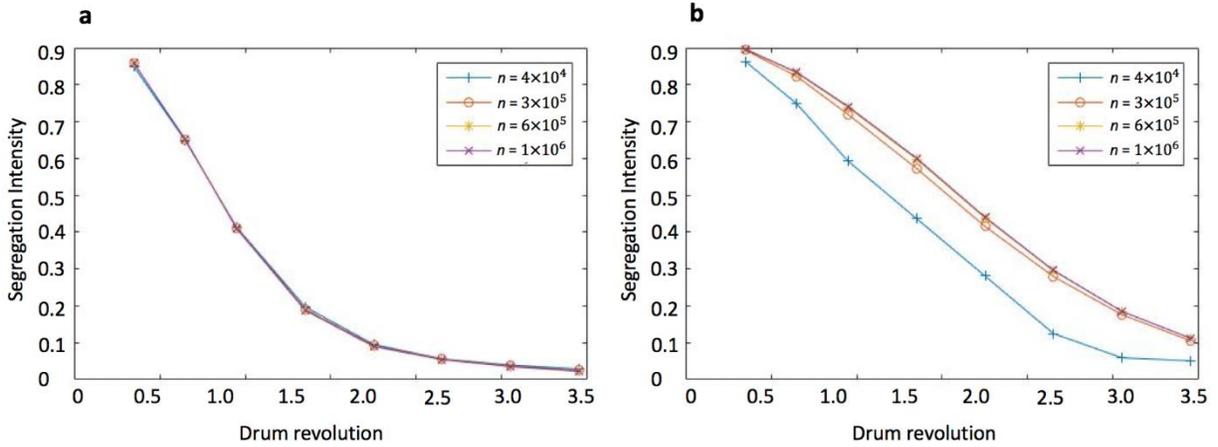

Figure 17. The segregation intensity for diffusion constants of (a) $k_1 = 0.04$ and (b) $k_1 = 0.005$.

To more efficiently investigate the influence of other parameters on the degree of mixing in the system (quantified using the segregation intensity), a sensitivity analysis of the system is performed first. Table 3 lists the dimensionless parameters resulting from the multi-scale model. For consistency, the Froude number $Fr$ and the filling volume fraction $f$ are kept constant throughout the study. Moreover, as mentioned previously, the elastic modulus $E$ and Poisson's ratio $\nu$ have little influence on the material flow behavior and, hence, are not studied. Therefore, the internal friction angle $f_{internal}$, wall friction angle $f_{wall}$, drum diameter to particle diameter ratio $D_{drum}/d$, and diffusion constants $k_1$ and $k_2$ are included in the parametric study.

Table 3. Dimensionless parameters.

| Parameter | Dimensionless Quantity |
| --- | --- |
| Segregation intensity | $I$ |
| Number of drum revolutions | $\dfrac{\omega t}{2\pi}$ |
| Froude number | $Fr = \dfrac{\omega^2 D_{drum}}{g}$ |
| Filling volume fraction | $f$ |
| Ratio of elastic modulus to max hydrostatic pressure | $\dfrac{E}{\rho_{bulk} g D_{drum} f}$ |
| Poisson's ratio | $\upsilon$ |
| Internal friction angle | $f_{internal}$ |
| Wall friction angle | $f_{wall}$ |
| Drum diameter to particle diameter ratio | $\dfrac{D_{drum}}{d}$ |
| Spanwise diffusion constant | $k_1$ |
| Streamwise diffusion constant | $k_2$ |

The effects of internal friction angle and wall friction angle are shown in Figures 18(a) and 18(b), respectively. A larger internal friction angle results in slower mixing, which appears to be due to a reduction in active region surface speeds caused by the increased frictional resistance. However, the mixing process happens so fast in this lab-scale drum that the materials are well-mixed after three revolutions, regardless of the internal friction angle. Figure 18(b) indicates that the mixing process is unaffected by the wall friction angle. The reason for this behavior is that, if the wall friction is sufficiently large to lift the powder, the first avalanche always occurs at the same location and the free surface angle remains constant (and equal to the internal friction angle). If the wall friction angle is too small, the powder cannot be lifted and it slips against the rotating wall.

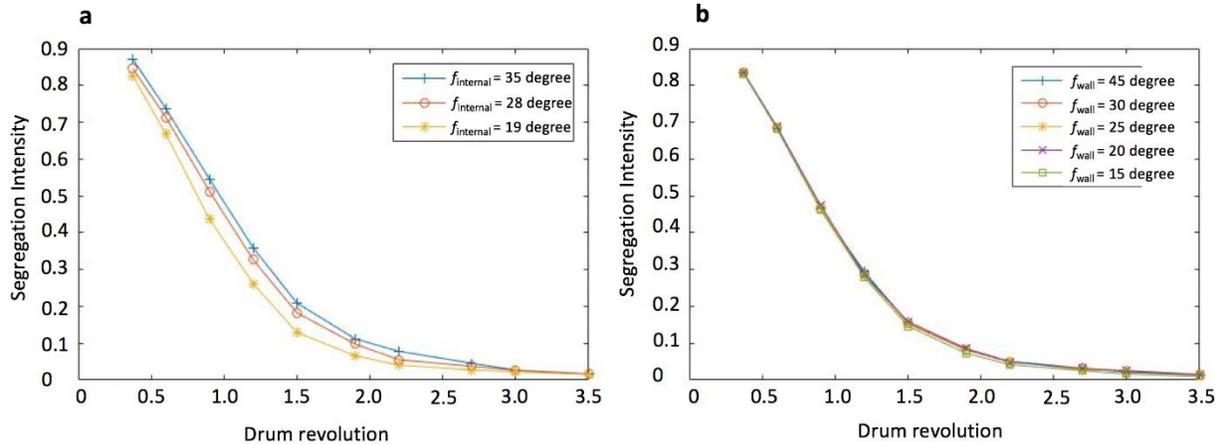

Figure 18. The segregation intensity *I* as a function of the number of drum revolutions for (a) different internal friction angles $f_{\text{internal}}$, and (b) different wall friction angles $f_{\text{wall}}$, as defined in Table 3.

The effect of drum-to-particle diameter ratio is shown in Figure 19. Although particles are not directly simulated in the multi-scale model, their size does indirectly appear in the calculation of the diffusion coefficient, with smaller particles resulting in smaller diffusion coefficients. The material properties remain unchanged since the material is assumed to remain cohesionless and identical in shape regardless of particle size. A larger drum-to-particle diameter ratio results in slower mixing since a larger drum diameter results in a thicker shear layer and a smaller particle diameter results in a smaller diffusion coefficient. The same trend can be found in works done by Kwapinska et al.[5] These results are useful for anticipating changes when scaling a mixing operation, for example.

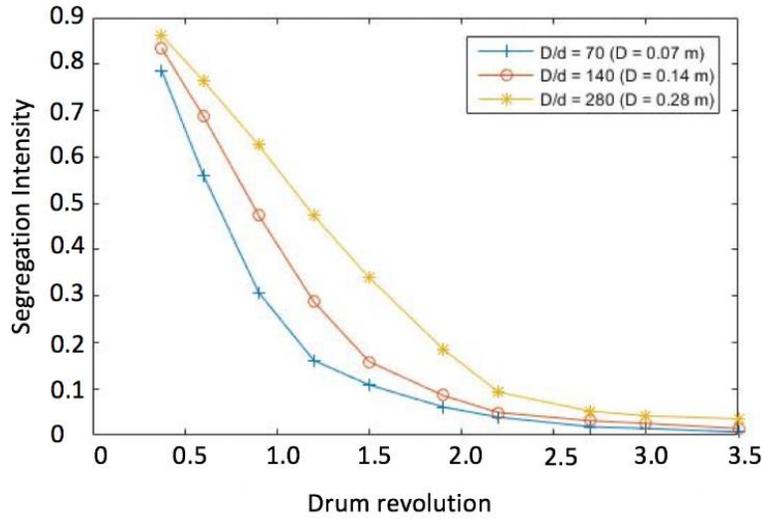

Figure 19. The segregation intensity as a function of the number of drum revolution for different drum diameter to particle diameter ratios $D_{\text{drum}}/d$, as defined in Table 3.

The effects of the spanwise and streamwise diffusion constants $k_1$ and $k_2$ are shown in Figures 20(a) and 20(b), respectively. According to Eq. (6), a larger diffusion constant $k_1$ or $k_2$ results in a larger diffusion coefficient and, thus, more rapid mixing. The diffusion constant $k_1$ is more dominant than $k_2$, even when $k_2$ is larger than $k_1$. In a rotating drum, as shown in Figure 11, the spanwise shear rate within the active region is much larger than the streamwise shear rate. Therefore, even though the streamwise diffusion constant $k_2$ is larger than the spanwise diffusion constant $k_1$, the mixing occurs mainly in the spanwise direction. However, this is not necessarily true for other geometries and hence the streamwise diffusion constant $k_2$ should be included in the diffusion coefficient expression for completeness.

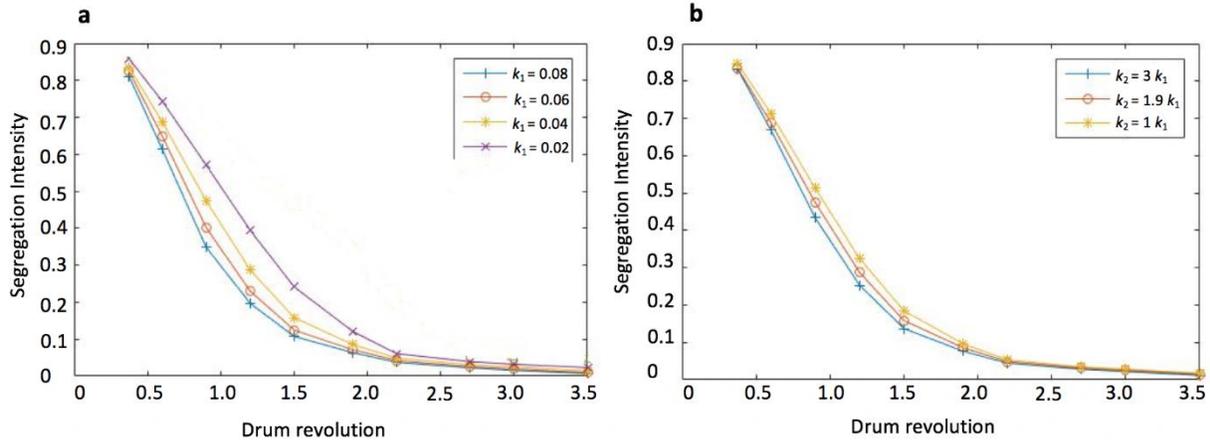

Figure 20. The segregation intensity as a function of the number of drum revolutions for (a) different spanwise diffusion constants $k_1$, and (b) different streamwise diffusion constants $k_2$, as defined in Table 3.

**Conclusions**

In this work, a new multi-scale approach to modeling particulate blending processes is presented. This multi-scale modeling approach combines finite element method simulations to obtain macroscopic velocity fields with calculations from the advection-diffusion equation with computationally and experimentally obtained expressions for particle diffusion at a local scale. The potential of this approach is demonstrated with the study of a rotating drum blender, i.e., with a "benchmark" system. However, the methodology proposed for obtaining the velocity field using FEM can also be used for more complex three-dimensional geometries and transient velocity fields., e.g., a Tote blender or a V blender. Predictions of concentration profiles and segregation intensity from the multi-scale model compare well quantitatively to DEM results, although the multi-scale model does predict smaller segregation intensities than those found from the DEM model at large times. This inaccuracy is due to the fact that the multi-scale model assumes continuum material behavior and, thus, the asymptotic mixing state corresponds to a perfectly mixed system as opposed to the asymptotic randomly mixed state predicted for the finite sized particles used in DEM simulations.

A significant advantage of the multi-scale blending model over DEM is that the multi-scale model is much faster to calculate. For the case examined here, corresponding to approximately 54,000 particles in the DEM simulation, the multi-scale model required approximately four to five hours of computation while the DEM model required five to six days. The time differences are expected to increase for larger systems since the number of DEM particles increases with the cube of the ratio of the system size to particle size while the FEM nodes increase, at most, only linearly with system size. Furthermore, if particle size is reduced in the DEM simulations, then the integration time step also decreases, further increasing the time required to complete DEM simulations. Hence, the real power of the multi-scale modeling approach is its ability to model industrially-relevant system sizes.

A second advantage of the multi-scale model over DEM is that all of the parameters used in the multi-scale model were measured from independent, standard tests or obtained from the literature. No back-fitting of the data was used to achieve good modeling accuracy. DEM models require a more complex approach to obtaining particle-level properties, such as complicated individual particle measurements, particularly for particles smaller than 1 mm, or time-consuming calibration simulations.

Parametric studies were performed using the multi-scale model to investigate the influence of various parameters on mixing behavior. First, the domain mesh size must be increasingly refined as the diffusion constants $k_1$ and $k_2$ decrease in order to minimize the influence of numerical diffusion. Second, increasing the internal friction angle decreases the mixing rate due to the decrease in surface speed. Third, the wall friction angle has little influence as long as the material is in an avalanching mode and not slipping against the wall since the free surface remains at the internal friction angle. Fourth, increasing the drum-to-particle diameter ratio decreases the rate at which overall mixing occurs. Finally, increasing the streamwise and spanwise diffusion constants increases the mixing rate; however, the spanwise diffusion constant dominates in a rotating drum due to the significant spanwise velocity gradient in the active zone.

The model presented here still has significant room for improvement. Future work should focus on modeling three-dimensional and unsteady flows, which are common in industrial practice. The effects of segregation could also be incorporated, as has been recently presented by Bertuola et al.[38] Of course, experimental validation at an industrial scale will be of critical importance in order to conclusively demonstrate the accuracy of this new modeling approach.